\documentclass[12pt]{iopart}

\usepackage{iopams}
\usepackage{cite}
\expandafter\let\csname equation*\endcsname\relax
\expandafter\let\csname endequation*\endcsname\relax
\usepackage[T1]{fontenc}

\usepackage{amsmath}
\usepackage{commath}
\usepackage{iopams}
\usepackage{graphicx}
\usepackage[breaklinks=true,colorlinks=true,linkcolor=blue,urlcolor=blue,citecolor=blue]{hyperref}
\usepackage{color}
\usepackage{mathrsfs}
\usepackage{longtable}

\def\P{\mathbb{P}}

\usepackage{graphicx}
\begin{document}

\title[]{Developing an Agent-Based Mathematical Model for Simulating Post-Irradiation Cellular Response: A Crucial Component of a Digital Twin Framework for Personalized Radiation Treatment}

\author{Ruirui Liu$^1$, Marciek H. Swat$^{2}$, James A. Glazier$^{3}$, Yu Lei$^{1}$, Sumin Zhou$^1$, Kathryn A. Higley$^{4}$}
\address{$^1$ Department of Radiation Oncology, University of Nebraska Medical Center, Omaha, Nebraska, USA}
\address{$^2$ Boulder BioConsulting, Inc., Boulder, Colorado, USA}
\address{$^3$ Biocomplexity Institute, Indiana University, Bloomington, Indiana, USA}
\address{$^4$ School of Nuclear Science and Engineering, Oregon State University, Corvallis, OR 97331, USA}

\begin{abstract}
In this study, we present the Physical–Bio Translator, a mechanistic, agent-based simulation framework for modeling radiation-induced cellular responses. The model incorporates a stochastic cell-state transition formalism that represents healthy, arrested, and dead phenotypes and accounts for both direct radiation damage and bystander signaling. We apply the framework to simulate monolayer cell cultures exposed to electron irradiation and analyze cell-cycle phase distributions, cell-state dynamics, and survival fractions under different dose and bystander conditions. The simulations qualitatively reproduce several expected features of cellular radiation response, including dose-dependent loss of proliferative capacity and modulation by indirect effects. These results demonstrate the feasibility of using a state-based mechanistic model within an agent-based environment as a building block for multiscale radiation-therapy digital twins. With additional experimental validation and parameter refinement, the Physical–Bio Translator could support in silico studies of radiation effects in cellular populations and contribute to more biologically informed radiation treatment modeling.
\end{abstract}

{\bf Keywords:} Cell State Transition Theory, Radiation-induced Cellular Effects, Radiation-induced Bystander Effect, Cell State Model, Cell Cycle Arrest, Digital Twin

\section{Introduction}

Optimizing radiation therapy remains a major challenge in oncology. Despite advances in devices, imaging, and dose planning, the complexity of biological responses—spanning DNA damage repair, cellular signaling, and tissue-level heterogeneity—still limits personalized treatment. Digital twins—dynamic virtual replicas integrating imaging, biological, and treatment data—offer a path forward by simulating tumor and tissue responses and guiding adaptive protocols \cite{R. Laubenbacher ,J. Jensen }. Realizing such clinically meaningful models, however, requires tackling the multiscale complexity of radiation biology, which current approaches cannot yet resolve.

Achieving a clinically meaningful digital twin for radiation therapy requires capturing radiation effects across vast spatial and temporal scales---from femtosecond atomic interactions to months of patient responses, and from nanometer DNA lesions to organ-level tumor control involving trillions of cells. Radiation response unfolds in stages: the \textit{physical stage} ($10^{-15}$--$10^{-9}$~s) of ionization and excitation; the \textit{chemical stage} ($10^{-9}$~s--1~s) of water radiolysis and radical reactions; and the \textit{biological stage} (seconds to months), encompassing DNA repair (NHEJ, HR, BER)~\cite{29}, checkpoint signaling, cell fate outcomes (apoptosis, senescence, autophagy, mitotic catastrophe), tissue-level changes (bystander effects, immune modulation, microenvironment remodeling), and tumor responses governed by the ``5~R's'' of radiobiology~\cite{ref4,ref5,ref6}. These cascades are further shaped by nonlinear feedback, homeostatic buffering, and patient-specific factors (e.g., age, genetics, comorbidities, tumor site, prior radiation, concurrent therapies). Such complexity overwhelms traditional modeling approaches, including purely mathematical analytical models, purely physics-based dose calculations, and simplified radiobiological formalisms such as the linear--quadratic survival model, making reliable prediction of treatment outcomes exceedingly difficult.

To realize clinically meaningful radiation digital twins, multiscale modeling is essential for integrating temporal and spatial scales into a coherent, dynamic, and mechanistic framework that captures both the biological cascade and patient-specific variability. A robust digital twin must incorporate models spanning the subcellular, cellular, multicellular, and tissue scales to comprehensively characterize and predict patient responses to radiation therapy. Such a multiscale approach enables a more detailed and accurate representation of the underlying biological processes, ultimately supporting more effective and individualized radiation treatment strategies.

Within this vision, each spatial scale requires appropriately designed mathematical models. A critical component is the cell-scale mechanistic model that simulates radiation-induced cellular effects. These effects originate from the initial physical interactions of ionizing radiation with cellular structures, as all subsequent biological responses are downstream consequences of early energy deposition events. Following irradiation, cells undergo a highly complex sequence of biochemical and signaling processes; however, given the current state of computational modeling, it is neither practical nor necessary to simulate every molecular event in detail. Instead, it is more effective to treat these processes from a mechanistic systems perspective—using the physical energy-deposition characteristics as inputs to a cellular-response model and representing the possible phenotypic outcomes (e.g., survival, apoptosis, senescence) as system-level outputs. This abstraction enables tractable simulation of cellular radiation responses while preserving key mechanistic relationships required for integration into a multiscale digital twin framework.

Historically, mechanistic models have been employed to simulate post-irradiation cellular responses, with numerous models proposed \cite{1,2,3,4,5}. These models have made significant contributions to radiation biology and radiation therapy. However, several challenges remain. For instance, emerging phenomena such as non-targeted effects have been increasingly recognized, yet most conventional modeling frameworks continue to rely on target-theory-based formulations. A generalized mechanistic model capable of capturing both targeted and non-targeted effects remains to be developed. Moreover, many existing models depend on simplified or phenomenological assumptions to represent processes that are not yet fully understood, resulting in limited mechanistic linkage between radiation dose and its microscopic biological outcomes at the cellular and tissue scales.

In response to the limitations of existing radiation effect models, we propose a new framework guided by both mechanistic and quantitative principles. A well-designed model should satisfy the following criteria:

\begin{itemize}
    \item It should be grounded in literature-supported, mathematically rigorous formulations that unify both target and non-target radiation effects within a single framework, while maintaining consistency with experimentally measured data.
    \item In keeping with the principle of parsimony, only essential biological and physical components should be included—each justified by prior evidence and contributing meaningfully to the model’s predictive capability.
    \item Model parameters should possess explicit biophysical meaning and, where possible, be constrained or estimated from measurable data rather than arbitrary fitting.
    \item The model should aim to provide mechanistic insights into radiation response while incorporating quantification principles common in physical sciences to enhance interpretability and reproducibility.
    \item Population-based parameterization should represent generalized biological behavior (e.g., an ``average'' cell or tissue response), improving the model’s transferability across experimental conditions.
    \item Finally, a modular design that integrates the physical, chemical, and biological stages of radiation interaction will facilitate extension, validation, and multi-scale simulation.
\end{itemize}

These criteria are not intended to claim that the present model fully meets all of these ideals. Rather, they serve as guiding principles for the broader goal of developing improved mechanistic models of radiation effects. The current work represents a step in this direction—demonstrating the feasibility of integrating essential mechanistic and quantitative elements within a unified modeling framework. Continued refinement and validation will further align the model with these principles as additional biological data become available.

In this study we propose a novel agent-based mathematical model that simulates the post-irradiation cellular response. From a metaphorical perspective, the mechanistic model for interpreting the radiation response after irradiation functions as a “translator” between physical interactions and biological interactions. So, we name our model as Physical-Bio Translator. In the current development of the Physical-Bio Translator, several major functions have been implemented, including simulating the bystander effect on cells, and predicting the possible phenotypes after irradiation. In this paper, we will present Physical-Bio Translator in detail.

\section{Methods and Materials}
\label{sec:2}
In this section, we introduce all the components of our developed agent-based mathematical model for simulating the radiation-induced cellular response.

\subsection{Cell-State Model}

To quantify the possible cellular phenotypes following irradiation, we introduce a cell state model. The central idea is that each observable post-irradiation phenotype corresponds to a specific underlying cell state; therefore, the distribution of phenotypes can be inferred by analyzing the distribution of cell states. The use of cell state–based representations in radiation modeling has a long history. For example, Albright proposed a cell state model to describe the time evolution of lesion burden after irradiation, classifying cell states into survival, uncommitted, and lethal categories \cite{6}. Little et al. applied a similar framework to simulate radiation-induced bystander effects, with cell states defined as alive but unaffected, affected and signaling, affected and non-signaling, or dead \cite{7}. Faria et al. also employed a cell state approach in modeling bystander effects, classifying cells simply as healthy or dead \cite{8}. These prior works demonstrate that cell state models provide a flexible and effective framework for linking radiation-induced physical and biochemical changes to observable cellular outcomes. More importantly, the cell state formalism offers a useful methodological foundation for quantifying the dynamic transition processes that cells undergo after irradiation—capturing how initial physical damage evolves through intermediate biological responses and ultimately manifests as distinct cellular phenotypes. This perspective aligns well with our goal of modeling radiation-induced cellular effects in a mechanistic yet tractable manner. Similarly, our cell-state model quantifies the possible cell phenotypes transition in temporal after irradiation.

\subsubsection{Cell State Classification} \hfill\\
We define three major cell states: \textbf{\textit {Healthy}}, \textbf{\textit{Arrested}}, and \textbf{\textit{Dead}}. A Healthy cell maintains its basic proliferation potential with no or very light damage. An Arrested cell has its cycle halted in a specific cell-cycle phase. A Dead cell has suffered irreparable damage and suspends material exchange with the extracellular matrix ({\it ECM}). We assume that the dead state is an average phenomenon of all the possible cell death routes, such as apoptosis, necrosis, etc. Each of these cell states differs depending on the cell’s phase in the cell cycle: G$_1$, S, G$_2$, or M.

\subsubsection{Cell State Transitions} \hfill\\
The allowed state transitions are:

\begin{itemize}
    \item From \textbf{\textit{Healthy}} to \textbf{\textit{Arrested}} or \textbf{\textit{Dead}}.
    \item From \textbf{\textit{Arrested}} to \textbf{\textit{Dead}} or \textbf{\textit{Healthy}}.
\end{itemize}

Dead cells stay dead. Transitions depend only on a cell’s current state. These state transition rules are based on radiation biology experiments. For instance, 1) A high radiation-induced damage causing direct cell death corresponds to the state transition from Healthy to Dead; 2) A moderate radiation-induced damage causing cell-cycle arrest corresponds to the state transition from Healthy to Arrested; 3) Cell apoptosis after failed cell damage repair during cell-cycle arrest, corresponds to the state transition from Arrested to Dead; 4) A dead cell does not have capacity to repair damage, so the Dead state is persistent. Because the dead state is an average phenomenon of different types of death, this formalism could be extended to include other definitions of cellular death types, e.g., cell apoptosis and necrosis.

For each cell state, we propose that there is a \textbf{\textit{state energy}} corresponding to the cell state, and the state energy is used to quantify the radiation-induced damage to each cell. The state energy includes two parts: the {\it direct state energy} $E_d$, which quantifies direct radiation-induced damage:
\begin{align}
\label{1}
E_d = \alpha N
\end{align} 
where $\alpha$ is a constant and $N$ is the number of double-strand DNA breaks (DSBs) produced by direct radiation hitting the cell. The DSB is the most lethal DNA damage type to the cell, and here we aggregate all direct damage into an effective DSB number.

The {\it indirect state energy} $E_i$ quantifies the net damage from bystander signaling. The bystander signals emitted by a signaling cell can migrate and interact with other cells.  In this work, we assume that the diffusion-reaction is the major process that contributes to the cell communication by bystander signals, and the receptor-ligand kinetics was considered to model the bystander signal and cell reaction process \cite{9}. We describe the interaction by a second-order reaction depending on the bystander signal concentration and cell receptor concentration:
\begin{align}
\label{2}
E_i = \beta \int^{t}_0\gamma(t')C_r(t')C_b(t')dt' 
\end{align} 
where $\beta$ is a constant, $t$ is the time after cell irradiation, $\gamma(t')$ is reaction rate coefficient at time $t'$, $C_r(t')$ is the transient concentration of cell receptor concentration at time $t'$, and $C_b(t')$ is the transient bystander signal concentration at $t'$.

Equation \eqref{2} integrates the absorption of bystander signals by the cell. For simplicity of denotation, we denote the integral of reaction rate as $C$, i.e.
\begin{align}
\label{3}
C =  \int^{t}_0\gamma(t')C_r(t')C_b(t')dt' 
\end{align} 

To be parallel to equation \eqref{1}, we can write the indirect state energy as:
\begin{align}
\label{4}
E_i = \beta C
\end{align}
where $C$ is the bystander signal concentration absorbed in the cell.
Both targeted and non-targeted cellular effect can lead to different level of DNA damage \cite{10}. In this work, we propose that state energy is a measure of DNA damage induced by radiation including target effect and the bystander effect. State energy is dimensionless, and two types of state energy could be added together, then the total state energy of a cell could be as:
\begin{align}
\label{5}
E = \alpha N + \beta C  
\end{align} 

When a cell experiences a radiation-induced damage directly or by absorbing bystander signals, its state energy increases. We define  $\Delta E$ as the  net increase of the state energy.

For a given level of radiation-induced damage, cells of the same type—which are otherwise indistinguishable—can exhibit markedly different outcomes: some remain healthy, others undergo cell-cycle arrest, and still others proceed to cell death. In other words, a fixed level of damage corresponds not to a single deterministic outcome but to a distribution of possible cell states, which in our framework maps to a distribution of cell state energies. Biologically, this implies that it is often impossible to assign a precise damage threshold to a specific post-irradiation phenotype. For example, when determining the lethal dose for radiation-induced cell death, experimental observations consistently show not a single lethal dose value but rather a range of doses over which cell death occurs. The same amount of radiation—or even the same level of bystander signaling—can induce different levels of damage across nominally identical cells, reflecting the inherently stochastic nature of radiation–cell interactions.

\begin{figure}[h]
\centering
\includegraphics[ width=0.8\textwidth]{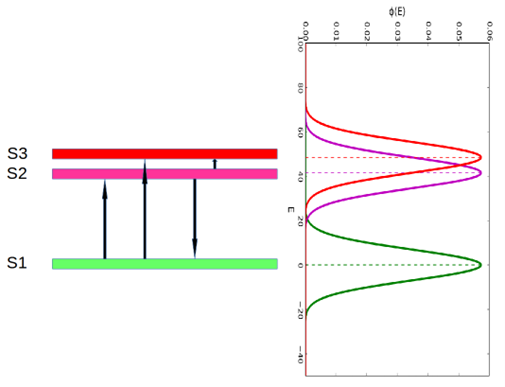}
\caption{The proposed three cell states, $S_1$, $S_2$ and $S_3$. The green bar corresponds to the healthy state, the pink one corresponds to the arrested state, and the red bar corresponds to the dead state. The black arrows indicate the possible cell state transition routes. Each cell state has a corresponding cell state energy distribution which is proposed as a Gaussian distribution.  
 }
\label{fig:1} 
\end{figure}

Here, we denote $S_1$, $S_2$, and $S_3$, representing the healthy state, arrested state and death state, respectively.  Here we propose that for each cell state $S_i$  $(i=1,2,3)$ its state energy E follows a normal distribution $N(E_i,\sigma^2_i)$ . The state energy $\phi_i(E)$ distribution of state $S_i$ could be written as:
\begin{align}
\label{6}
\phi_i(E) = \frac{1}{\sqrt{2\pi}\sigma_i} \exp\left(- \frac{(E-E_i)^2}{2\sigma^2_i} \right)
\end{align} 
where $E_i$ is the mean state energy of state $S_i$, and  is the standard deviation of state energy of state $S_i$. One example of a state energy distribution is shown in Figure\ref{fig:1}. 

When radiation-induced DNA damage is incurred, the cell state energy will increase, which will induce a change in the cell state energy distribution correspondingly. We assume that the shape of the state energy distribution will not change, but the mean state energy will shift. For instance, suppose the current energy state of the cell is $S_i$, with the net increase of state energy $\Delta E$, its state energy distribution will be as:
\begin{align}
\label{7}
\phi_i(E) = \frac{1}{\sqrt{2\pi}\sigma_i} \exp\left(- \frac{(E- (E_i+\Delta E))^2}{2\sigma^2_i} \right)
\end{align} 

Given  the state energy distribution, it is plausible to quantify the probability of state transitions. The cell state transition probability is quantified by calculating the overlapping integral of the state energy distribution. 

The overlapping integral of two state energy distributions is written as:
\begin{align}
\label{8}
\left\langle \phi_i \vert \phi_j \right\rangle = \int_D \min [\phi_i(x), \phi_j(x)] dx
\end{align} 
where $\phi_i(x)$ and $\phi_j(x)$ are the state energy distribution functions of two cell energy states. 

The computation of the overlapping integral for two state energy distributions, $N(E_i, \sigma_i^2), N(E_j, \sigma_j^2)$, depends on whether the two variances are equal or not. Here we start with the simpler case where we have $\sigma_1^2=\sigma_2^2=\sigma^2$. We then have 
\begin{align}
\label{9}
\left \langle \phi_i \vert \phi_j \right\rangle = 2\Phi \left( - \frac{\lvert E_i-E_j\rvert}{2\sigma} \right) 
\end{align} 
where $\Phi(x)$ is the cumulative distribution function of the normal distribution and it is as 
\begin{align*}
\Phi(x) = \int^x_{-\infty}\frac{1}{\sqrt{2\pi}} \exp\left(- \frac{t^2}{2} \right) dt
\end{align*} 

It is obvious that $0 \leq \left< \phi_i \vert \phi_j \right> \leq 1$, and $\left< \phi_i \vert \phi_j \right> =1$ if and only if those two normal distributions are identical. 
We use the overlapping integral as a measure of cell state transition probability between two states, and the transition probability from state $S_i$ to $S_j$ is $P(S_i \rightarrow S_j)$, and
\begin{align}
\label{10}
P(S_i \rightarrow S_j) = \left< \phi_i \vert \phi_j \right> .
\end{align} 

Suppose initially at time $t$, a cell stays at state $S_i$. Given the increase of state energy $\Delta E$, the cell will shift to a possible state $S_j$, and the transition probability  is
\begin{align}
\label{11}
P(S_i \rightarrow S_j; \Delta E) = 2\Phi\left( -\frac{\vert E_i+\Delta E-E_j \vert}{2\sigma}\right) 
\end{align}

We note that, biologically, the cell state energy should not take negative values. In our formulation, however, $E$ is treated as a latent continuous variable defined on $\mathbb{R}$, and the normal distribution in Eq.~\eqref{6} is used as a convenient mathematical approximation to describe the variability of state energy around its mean. For the parameter ranges considered in this work, the means $E_i$ are sufficiently larger than the corresponding standard deviations $\sigma_i$ such that the probability mass in the non-physical region $E<0$ is negligible and does not materially affect the calculated transition probabilities. More formally, one could employ a truncated or log-normal distribution with support on $E \ge 0$, but the Gaussian assumption yields closed-form expressions for the overlap integrals and keeps the model analytically tractable while remaining a good approximation in the biologically relevant energy range.

Equation~\eqref{11} shows that the transition probability between two states depends on the overlap between their state energy distributions. For transitions into the lethal state $S_3$ (i.e., $S_1 \rightarrow S_3$ or $S_2 \rightarrow S_3$), the overlap increases as the shifted mean $E_i + \Delta E$ approaches $E_3$, but then decreases again if the mean moves far beyond $E_3$. Biologically, however, once the accumulated radiation damage becomes extremely large, the probability of cell death should not decrease; rather, it should approach a saturation value close to unity.

Because this non-monotonic behavior is specific to transitions into the lethal state $S_3$, we introduce a saturation boundary condition only for $S_i \rightarrow S_3$ transitions. Let
\[
\Delta E_{\text{sat}} = \max(0, E_3 - E_i)
\]
denote the net increase in state energy needed for the mean of the current state to reach that of the lethal state. The transition probability to $S_3$ is therefore defined as
\begin{align}
P(S_i \rightarrow S_3; \Delta E) =
\begin{cases}
2\Phi\!\left(-\dfrac{|E_i + \Delta E - E_3|}{2\sigma}\right), & 0 \le \Delta E \le \Delta E_{\text{sat}},\\[8pt]
1, & \Delta E > \Delta E_{\text{sat}}.
\end{cases}
\label{eq:satP}
\end{align}

This modification applies only to lethal-state transitions and leaves all other transitions (e.g., $S_1 \rightarrow S_2$ or $S_2 \rightarrow S_1$) unchanged. It preserves the overlap-based formulation in the small- to moderate-damage regime while enforcing biologically realistic saturation of cell-killing probability at high levels of radiation-induced damage.

\begin{figure}[h]
\centering
\includegraphics[width=.6\textwidth]{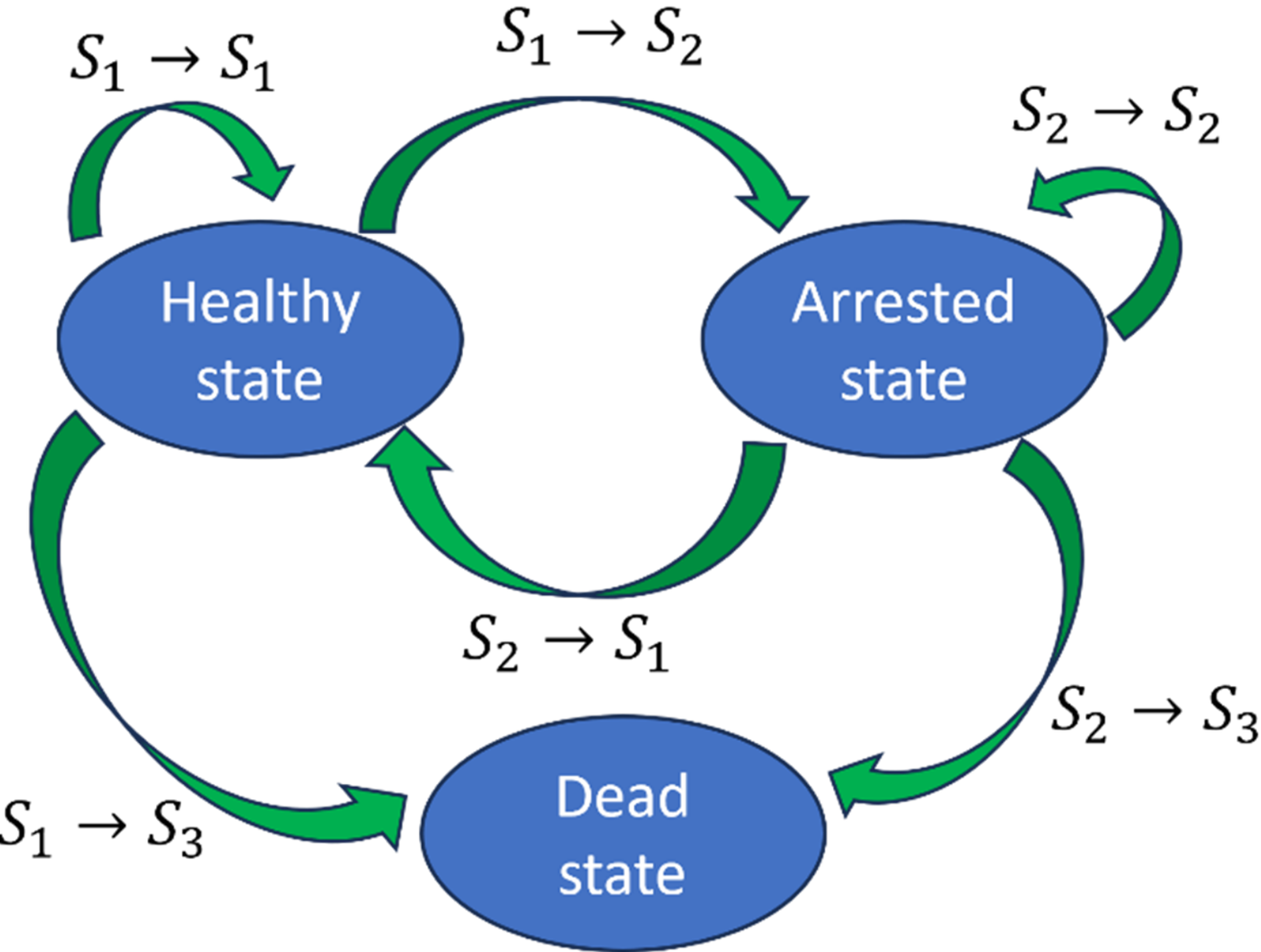}  
\caption{Cell state transition diagram}
\label{fig:2} 
\end{figure}

Based on equation \eqref{11}, we can calculate the probabilities for all the possible transition routes. The possible cell state transition route is shown in Figure \ref{fig:2}. The final transition route is selected based on a rejection algorithm based on Monte Carlo sampling \cite{11}. For instance, for quantifying the transition route for $S_1$, the selection process is as follows:
\begin{itemize}
    \item if $\displaystyle \xi \leq \frac{P(S_1 \rightarrow S_1)}{P(S_1 \rightarrow S_1)+P(S_1 \rightarrow S_2)+P(S_1 \rightarrow S_3)}$, where $0< \xi < 1$, the cell remains in $S_1$
    \item  if $\displaystyle \frac{P(S_1 \rightarrow S_1)}{P(S_1 \rightarrow S_1)+P(S_1 \rightarrow S_2)+P(S_1 \rightarrow S_3)} < \xi \leq \frac{P(S_1 \rightarrow S_2)}{P(S_1 \rightarrow S_1)+P(S_1 \rightarrow S_2)+P(S_1 \rightarrow S_3)}$, the cell transitions into $S_2$
    \item Otherwise, cell transitions into $S_3$
\end{itemize}

\subsubsection{Radiation Sensitivity in Different Cell Phases} \hfill\\
Cells are known to have different radiation sensitivity in different cell phases. Cells in the G$_2$ and M phases usually have higher radiation sensitivity, lower radiation sensitivity in the G$_1$ phase, and the lowest radiation sensitivity during the latter part of the S phase. We define a radiation sensitivity factor $f$ for each cell phase. Here, we introduce a method to calculate the radiation sensitivity factor using the cell state transition theory we just introduced above.

For nomenclature simplicity, each cell phase is assigned an index i corresponding to the four cell phases where $i \in \{1,2,3,4\}$ and $i=1$ corresponds to G$_1$, $ i=2$ corresponds to S, $i=3$ corresponds to G$_2$, and $i=4$ corresponds to M. We denote $E_{i, j}$ as the mean state energy of state $S_j$ at $i^{th}$ cell phase, i.e., $S_{i, j}$ where $i \in \{1,2,3,4\}$ and $j\in \{1,2,3\}$.
 
For cell phase G$_1$, the transition probability from $S_1$ to $S_3$ with net increase of state energy $\Delta E$ is
\begin{align}
\label{12}
P_1(S_1 \rightarrow S_3; \Delta E) = 2\Phi\left( -\frac{\vert E_{1, 1}+\Delta E - E_{1, 3} \vert}{2\sigma}\right) 
\end{align}

Now we consider a less radiosensitive cell phase, S, the transition probability from $S_1$ to $S_3$ with $\Delta E$ is 
\begin{align}
\label{13}
P_2(S_1 \rightarrow S_3; \Delta E) = 2\Phi\left( -\frac{\vert E_{2, 1}+\Delta E - E_{2, 3} \vert}{2\sigma}\right)
\end{align}

From equation \eqref{12} and equation \eqref{13} we can know that transition probability is a function of $\Delta E$  . Here, we calculate the derivatives of equation \eqref{12} and equation \eqref{13} with respect to $\Delta E$. We can have 
\begin{align}
\label{14}
\frac{dP_1(S_1 \rightarrow S_3; \Delta E)}{d\Delta E} = -\frac{1}{\sigma}\,g\,\left( -\frac{\vert E_{1, 1}+\Delta E - E_{1, 3} \vert}{2\sigma}\right)\frac{E_{1, 1}+\Delta E - E_{1, 3}}{\vert E_{1, 1}+\Delta E - E_{1, 3}\vert}
\end{align}

where $g=\frac{1}{\sqrt{2\pi}} \exp(- \frac{t^2}{2})$ which is the normal distribution density function. Similarly, we can get 

\begin{align}
\label{15}
\frac{dP_2(S_2 \rightarrow S_3; \Delta E)}{d\Delta E} = -\frac{1}{\sigma}\,g\,\left( -\frac{\vert E_{2, 1}+\Delta E - E_{2, 3} \vert}{2\sigma}\right)\frac{E_{2, 1}+\Delta E - E_{2, 3}}{\vert E_{2, 1}+\Delta E - E_{2, 3}\vert}
\end{align}

We define the radiation sensitivity factor $f_i$ as a relative measure of how strongly a small net increase of state energy  $\Delta E$ changes the transition probability from $S_1$ to $S_3$ in phase $i$. We take $f_1 = 1$ for the G$_1$ phase as a reference, and define the radiation sensitivity factor for the S phase as
\begin{align}
\label{eq:f2}
f_2 
= \lim_{\Delta E \to 0}
\frac{\dfrac{\mathrm{d}P_2(S_1 \rightarrow S_3; \Delta E)}{\mathrm{d}\Delta E}}
     {\dfrac{\mathrm{d}P_1(S_1 \rightarrow S_3;\Delta E)}{\mathrm{d}\Delta E}}.
\end{align}

Intuitively, for small $\Delta E$ we can approximate the transition probabilities by a first-order expansion,
\[
P_i(S_1 \rightarrow S_3; \Delta E) 
\approx P_i(S_1 \rightarrow S_3; 0) 
+ \left.\frac{\mathrm{d}P_i}{\mathrm{d}\Delta E}\right|_{\Delta E = 0}\!\Delta E,
\]
so the derivative $\left.\frac{\mathrm{d}P_i}{\mathrm{d}\Delta E}\right|_{\Delta E = 0}$ represents the incremental change in the probability of transitioning from $S_1$ to $S_3$ per unit net increase of state energy in phase $i$. The factor $f_2$ in Eq.~\eqref{eq:f2} is thus the ratio of these slopes between the S and G$_1$ phases. Values $f_2 < 1$ correspond to lower radiosensitivity in S relative to G$_1$, whereas $f_2 > 1$ would indicate higher radiosensitivity. In this way, $f_2$ provides a cell-cycle-phase--specific, mechanistically derived scaling of radiosensitivity, analogous to a relative sensitivity factor based on how efficiently net increase of state energy drives transitions to severely damaged states.

Then according to equation \eqref{14} and equation \eqref{15}, we can get 
\begin{align}
\label{17}
f_2 = \exp \left( -\frac{(E_{2, 1}-E_{2, 3})^2}{8\sigma^2} + \frac{(E_{1, 1}-E_{1, 3})^2}{8\sigma^2} \right)
\end{align}

Then we can have
\begin{align}
\label{18}
(E_{1,1}-E_{1,3})^2 - (E_{2,1}-E_{2,3})^2 = 8\sigma^2\ln f_2
\end{align}

Equation \eqref{18} shows the relationship between radiation sensitivity factor and the mean state energies. 

\subsubsection{Cell-Cycle Arrest Duration After Irradiation}\hfill\\
It is well established that cells may undergo temporary or permanent cell-cycle arrest following irradiation. Here, we propose a method to calculate the arrest duration using the cell-state transition model previously introduced.

Experimental studies have shown that DNA double-strand break (DSB) repair exhibits biphasic kinetics, with an initial fast component followed by a slower phase of rejoining~\cite{12,13}. A commonly used formulation is the two-component first-order repair model~\cite{14,15,16}, in which the total number of unrepaired DSBs at time $t$ is expressed as
\begin{align}
\label{19}
N(t) = f_{\mathrm{fast}} \, e^{-\lambda_1 t} + f_{\mathrm{slow}} \, e^{-\lambda_2 t},
\end{align}
where $f_{\mathrm{fast}}$ and $f_{\mathrm{slow}}$ are the fractional contributions of the fast and slow repair components, respectively (with $f_{\mathrm{fast}} = 1 - f_{\mathrm{slow}}$), and $\lambda_1$ and $\lambda_2$ are the corresponding rate constants.

\vspace{1ex}
\paragraph{Transient arrest and arrest duration.}
We define the \textit{transient arrest state} as a reversible state in which a cell temporarily halts its progression through the cell cycle. The \textit{transient arrest duration} is the total time a cell remains in this state before returning to normal cycling.

In the cell-state transition model, the state energy serves as a surrogate for the degree of radiation-induced damage. Motivated by the biphasic repair behavior of DSBs, we assume that the state energy of a cell in the arrested state $S_2$ decreases over time following the same kinetic pattern:
\begin{align}
\label{20}
E(t) = E_0 \big( f_{1} \, e^{-\lambda_1 t} + f_{2} \, e^{-\lambda_2 t} \big),
\end{align}
where $E_0$ is the initial state energy immediately after transition into $S_2$, and $f_1, f_2$ correspond to the fractional contributions of the two repair components.

\vspace{1ex}
\paragraph{Probability of transition from arrest to a healthy state.}
A cell exits the arrested state $S_2$ and returns to the healthy state $S_1$ once sufficient repair has occurred. The probability of transition at time $t$ is determined by the overlap between the instantaneous state-energy distribution of $S_2$ and that of $S_1$.  The transition probability is
\begin{align}
\label{21}
P(t) = 2\Phi\!\left( -\frac{|E_2(t) - E_1|}{2\sigma} \right),
\end{align}
where $\Phi(\cdot)$ is the standard normal cumulative distribution function and $E_2(t)$ is given by
\begin{align}
\label{22}
E_2(t) = E_2(0)\big(f_1 e^{-\lambda_1 t} + f_2 e^{-\lambda_2 t}\big).
\end{align}

The time derivative yields the transition probability rate:
\begin{align}
\label{24}
\frac{dP(t)}{dt}
= -\frac{1}{\sqrt{2\pi}\sigma}
\exp\!\left(-\frac{E_2^2(t)}{8\sigma^2}\right)
\frac{dE_2(t)}{dt}.
\end{align}

\vspace{1ex}
\paragraph{Mean arrest duration.}
The mean lifetime of the arrested state $S_2$, denoted $T_m$, is given by
\begin{align}
\label{25}
T_m = \int_0^{\infty} t\, \frac{dP(t)}{dt} \, dt.
\end{align}
Substituting Eq.~\eqref{22} and rearranging, we obtain
\begin{align}
\label{26}
T_m = \frac{E_2(0)}{\sqrt{2\pi}\sigma}\!
\int^{\infty}_0  
t\, \exp\!\left( -\frac{E_2^2(t)}{8\sigma^2} \right)
\big( f_1 \lambda_1 e^{-\lambda_1 t}
    + f_2 \lambda_2 e^{-\lambda_2 t} \big)\, dt.
\end{align}

To obtain a practical closed-form approximation, we expand the exponential term, $\exp(-\frac{E^2_2(t)}{8\sigma^2})$ and truncate higher-order contributions, yielding:
\begin{align}
\label{27}
T_m \approx
\frac{E_2(0)}{\sqrt{2\pi}\sigma}
\left( \frac{f_1}{\lambda_1} + \frac{f_2}{\lambda_2} \right).
\end{align}

Equation~\eqref{27} indicates that the slow repair component (with the smaller decay constant $\lambda_2$ and thus longer half-life) has the dominant influence on arrest duration. This behavior is consistent with experimental findings~\cite{17} showing that the slow homologous recombination repair pathway governs the characteristic timescale of radiation-induced cell-cycle arrest.

\subsection{Agent-based simulation}
In this study, we implement the proposed cell-state model within an agent-based simulation framework to represent the collective cellular response after irradiation. Our approach follows the conceptual structure of agent-based modeling—each cell is treated as a simple finite-state unit with well-defined update rules—while relying on the cell-state transition theory we developed to greatly reduce computational burden. Because the model abstracts cellular behavior into probabilistic state transitions rather than resolving detailed molecular or biophysical dynamics, simulations involving tens of thousands of cells, and potentially larger populations, can be performed efficiently on standard computational hardware. The framework consists of two components: the cellular lattice and the transition rules. Each cell is represented as a finite-state machine \cite{18}, indexed by $i$, with its state at time $t$ denoted by $S_i^t$. The neighborhood of cell $i$ is denoted by $\eta_i^t$, and each cell updates its state based on its own current state and the states of its neighbors. This stochastic, state-based formulation captures the essential population-level dynamics needed for modeling tissue-scale radiation response while maintaining computational tractability.

\begin{align}
\label{28}
S^{t+1}_i = F(S^t_i, \eta^t_i)
\end{align}
At each time step, all cells update their state synchronously according to $ F(S^t_i, \eta^t_i)$.

\subsubsection{Cell-Cycle Phase Duration and Transition Rules} \hfill\\
As soon as the duration of a given cell-cycle phase has elapsed, the cell transitions to the next phase. The phase duration is not fixed but instead varies stochastically according to a phase-specific probability density function. Let $f_i(t)$ denote the probability density of the duration of the $i^{\text{th}}$ cell-cycle phase. The probability that the duration lies in the interval $[t, t + \mathrm{d}t]$ is $f_i(t)\,\mathrm{d}t$. Following established formulations \cite{19}, we model $f_i(t)$ as a half-Gaussian distribution,
\begin{equation}
\label{29}
f_i(t) =
\begin{cases}
0, & t < T_i,\\[6pt]
\dfrac{1}{\sqrt{2\pi}\sigma_i}
\exp\!\left[-\dfrac{(t - \bar{t}_i)^2}{2\sigma_i^2}\right], & t \ge T_i,
\end{cases}
\end{equation}
where $\bar{t}_i$ and $\sigma_i^2$ are the mean and variance of the duration of phase $i$, respectively.

In simulations, the duration of each phase is sampled using Monte Carlo sampling. For the $i^{\text{th}}$ phase, a sample duration $T_i$ is drawn as
\begin{align}
\label{30}
T_i = \bar{t}_i + \sigma_i \sqrt{-2\ln \xi_1}\,
\cos\!\left(\frac{\pi}{2}\,\xi_2\right),
\end{align}
where $\xi_1$ and $\xi_2$ are independent random numbers uniformly distributed in $(0,1)$.

\paragraph{Indicator-based transition rules.}
To formalize the cell-phase transition mechanism, we define each rule as an
\emph{indicator function} that returns 1 if the transition condition is satisfied
and 0 otherwise.

\textbf{(i) Phase-duration rule:}
\[
F_{1}(S_i^t,\eta_i^t) =
\begin{cases}
1, & t > T_i,\\
0, & t \le T_i,
\end{cases}
\]
which represents whether the cell has spent sufficient time in its current phase.

\textbf{(ii) Neighbor-availability rule:}
Cell-cycle progression also depends on local conditions, including contact inhibition and available space for daughter cells. Following standard formulations \cite{20,21}, we define
\[
\eta_i^t = P(i-1,j) + P(i+1,j) + P(i,j-1) + P(i,j+1),
\]
where $P(i,j)=1$ if the lattice site $L(i,j)$ is empty and $P(i,j)=0$ otherwise, and $\eta_i^t \in \{0,1,2,3,4\}$. The neighbor-availability rule is then
\[
F_{2}(S_i^t,\eta_i^t) =
\begin{cases}
1, & \eta_i^t > 0,\\
0, & \eta_i^t = 0.
\end{cases}
\]

\textbf{(iii) Combined transition rule:}
The cell progresses to the next phase only if \emph{both} conditions are satisfied.  
We therefore define the combined rule as the product of the indicator functions:
\begin{align}
\label{34}
F(S_i^t,\eta_i^t) = F_{1}(S_i^t,\eta_i^t)\, F_{2}(S_i^t,\eta_i^t).
\end{align}
If $F = 1$, the cell advances to the next phase; if $F = 0$, the cell remains in its current phase. This multiplicative formulation provides a mathematically explicit and extensible mechanism for incorporating additional biological constraints when needed.

\paragraph{Cell-phase distribution.}
We track the cell-cycle distribution using the cell-phase ratio. Let $N_{P_i}$ denote the number of cells in phase $i$ and $N_P$ the total number of proliferating cells. The phase ratio is defined as
\begin{align}
\label{35}
f_{P_i} = \frac{N_{P_i}}{N_P}.
\end{align}

\begin{figure}[h]
\centering
\includegraphics[width=.6\textwidth]{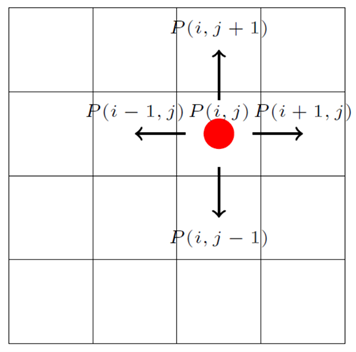}
\caption{Illustration of neighborhood-based space availability.
The current cell occupies $P(i,j)$; daughter cells can be placed in any of
the four adjacent vacant sites.}
\label{fig:3}
\end{figure}

\subsubsection{Simulating Cell State Transitions} \hfill\\
The evolution of cell states after irradiation is a continuous stochastic process.  
The transition probability obtained from Eq.~\eqref{11} represents the total probability
that a cell initially in state $S_i$ will eventually transition to state $S_j$ as a result of
radiation-induced perturbation.  
In an agent-based framework, however, time is discretized, and the transition probability
must be evaluated at each time step.  
To do this in a consistent manner, we introduce a stochastic time-to-transition model that 
converts the total transition probability into a per–time-step probability.

At each simulation time step, the net increase of state energy $\Delta E$ for each cell
is first computed from the contributing sources (e.g., direct DNA double-strand breaks,
bystander signals).  
Direct DSBs generally cause rapid and severe perturbations, whereas bystander signals induce
weaker but persistent stresses.  
To capture this distinction, we introduce two conceptual modes of transition:
\textit{instantaneous transitions}, triggered immediately when $\Delta E>0$, and
\textit{delayed transitions}, in which the transition occurs stochastically over time.

For delayed transitions, we model the state-change process as a failure-time problem.
Let $T$ denote the (random) time at which the transition occurs, and let $T_o$
represent the observation window within which radiation-induced changes can be measured.
The cumulative distribution function of the transition time is
\begin{align}
\label{36}
F(t) = P(T \le t), \qquad t \ge 0,
\end{align}
with corresponding density
\begin{align}
\label{37}
f(t) = \lim_{\Delta t \to 0} P(t < T \le t+\Delta t).
\end{align}
The total probability of transition within the observation window is
\begin{align}
\label{38}
p_o = \int_{0}^{T_o} f(t)\, dt.
\end{align}
For a transition from $S_i$ to $S_j$, the total transition probability is given by
Eq.~\eqref{11}:
\begin{align}
\label{39}
p_o = 2\Phi\!\left(-\frac{|E_i + \Delta E - E_j|}{2\sigma}\right).
\end{align}

To convert $p_o$ into a time-resolved probability, we assume that $T$ follows an
exponential distribution with rate parameter $\lambda$,
\begin{align}
\label{40}
f(t) = \lambda e^{-\lambda t}.
\end{align}
Matching Eq.~\eqref{38} with Eq.~\eqref{39} yields
\begin{align}
\label{41}
\int_0^{T_o} \lambda e^{-\lambda t}\, dt = p_o
\quad \Rightarrow \quad
\lambda = -\frac{\ln(1 - p_o)}{T_o}.
\end{align}

\paragraph{Per–time-step transition probability.}
Without loss of generality, consider a simulation time step $[t,\, t+\Delta t]$.
The conditional probability that the cell completes the state transition within
this interval, given that the transition has not yet occurred by time $t$, is
\begin{align}
\label{43}
p(t < t' \le t+\Delta t \mid T > t)
= \frac{P(t < T \le t+\Delta t)}{P(T > t)},
\end{align}
where $t'$ denotes the (random) time at which the cell completes the state transition
within the interval.  
Since $T$ is exponentially distributed, we obtain
\begin{align}
\label{44}
p(t < t' \le t+\Delta t \mid T > t)
= \frac{\int_{t}^{\,t+\Delta t} \lambda e^{-\lambda x}\, dx}
       {1 - \int_{0}^{t} \lambda e^{-\lambda x}\, dx}
= 1 - e^{-\lambda \Delta t}.
\end{align}

Thus, the probability of transition in each time step is
\[
p_{\Delta t} = 1 - e^{-\lambda \Delta t}.
\]

The time step $\Delta t$ is chosen to balance computational cost and temporal resolution;
in this work, we use $\Delta t = 1$ minute.

\paragraph{Indicator-based transition rule.}
To implement the stochastic transition in the simulation, we define the transition rule
as an indicator function:
\begin{align}
\label{45}
F_{\mathrm{trans}}(\lambda,\Delta t,\xi) =
\begin{cases}
1, & \xi < p_{\Delta t},\\
0, & \xi \ge p_{\Delta t},
\end{cases}
\end{align}
where $\xi \sim U(0,1)$ is a uniformly distributed random number.
If $F_{\mathrm{trans}} = 1$, the cell undergoes the state transition; if $F_{\mathrm{trans}} = 0$,
the cell remains in its current state.
This indicator formulation is consistent with the transition-rule methodology introduced
earlier and can be readily combined multiplicatively with other rule components when
multiple conditions govern state changes.

\paragraph{Cell-state distribution.}
The overall state composition of the simulated population is tracked using the state ratio:
\begin{align}
\label{46}
f_{S_i} = \frac{N_{S_i}}{N_S},
\end{align}
where $N_{S_i}$ is the number of cells in state $S_i$ and $N_S$ is the total number of cells.

\subsection{Simulation example}

\begin{figure}[h]
\centering
\includegraphics[width=.85\textwidth]{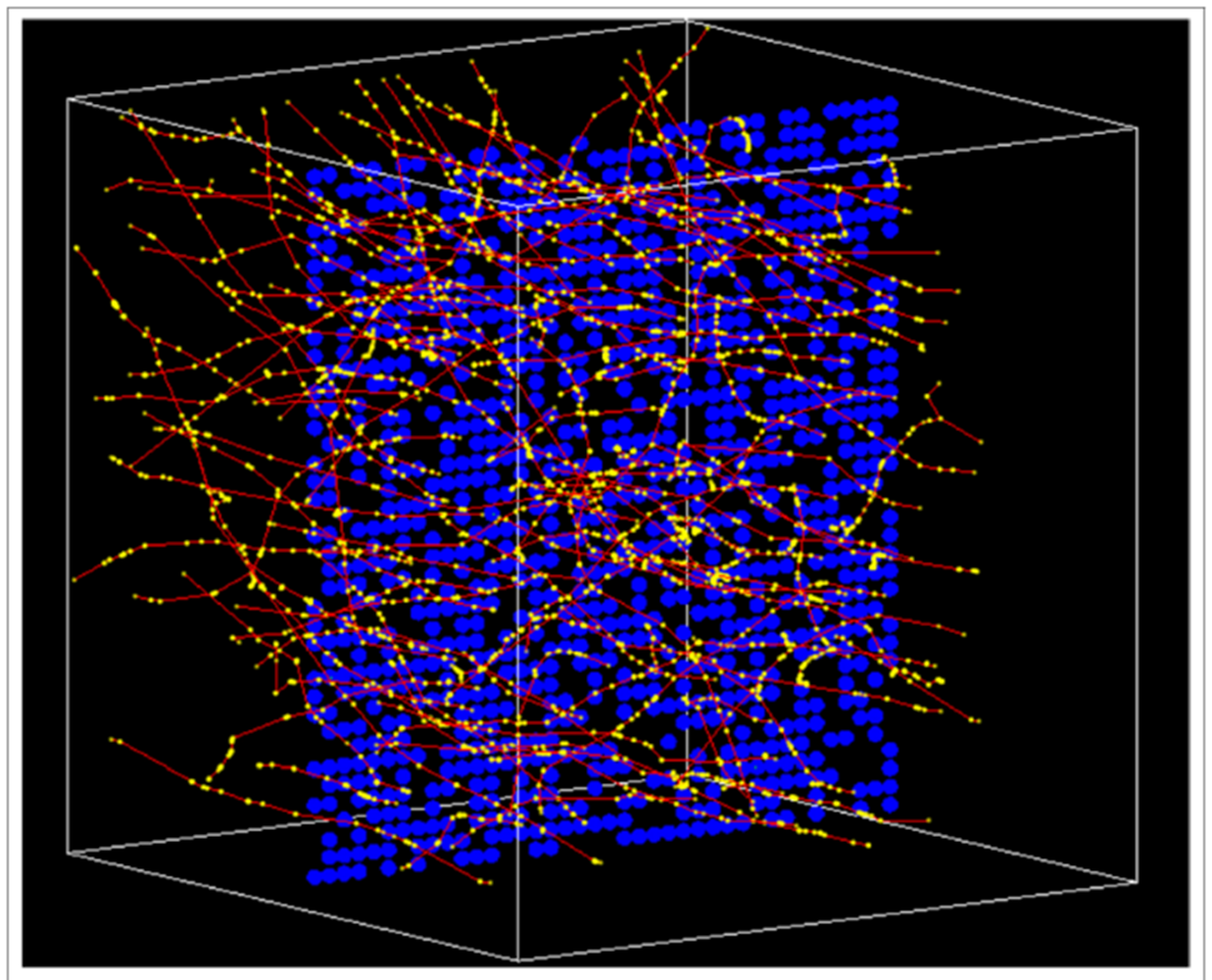}
\caption{Monolayer cell culture irradiated by a 1~MeV electron plane source.
Each cell is modeled as a 40~\textmu m water sphere with a 10~\textmu m nucleus
centered inside. The red lines with yellow points illustrate representative
electron tracks generated by the Geant4-DNA physics processes.}
\label{fig:4}
\end{figure}

To demonstrate the application of the proposed framework, we simulate radiation
response in a monolayer cell culture composed of 1000 cells irradiated by a
1~MeV electron plane source.  
The irradiation geometry and cell layout are shown in Fig.~\ref{fig:4}.

\paragraph{Direct radiation effects (Geant4-DNA).}
The irradiation simulation is performed using the Geant4 toolkit \cite{22}
with the Geant4-DNA physics models enabled for track-structure simulations in
liquid water.  
Geant4-DNA provides discrete electron–water interaction processes, allowing us
to compute the spatial distribution of energy-deposition events within each cell
at nanometer resolution.  
From these recorded energy-deposition points, the absorbed dose and the number
of DNA double-strand breaks (DSBs) in each cell are calculated.  
DSBs are identified using a DBSCAN clustering algorithm
\cite{23,24,25}, where individual energy-deposition events are grouped into
clusters if their spatial separation is below a predefined interaction radius.
This approach is consistent with commonly used geometric clustering methods for
classifying DSB-producing events in track-structure simulations.

\paragraph{Indirect radiation effects (bystander signaling).}
To model the contribution of radiation-induced bystander signaling, we solve a
two-dimensional reaction–diffusion equation over the monolayer domain:
\[
\frac{\partial C(x,y,t)}{\partial t}
= D\nabla^2 C(x,y,t) - \kappa C(x,y,t) + S(x,y,t),
\]
where $C(x,y,t)$ denotes the concentration of bystander signal, $D$ is the
diffusion coefficient, $\kappa$ is the first-order decay constant, and
$S(x,y,t)$ represents signal production by irradiated cells.
This formulation follows standard bystander-signal modeling approaches
\cite{26,27}.  
The equation is solved numerically on a spatial grid aligned with the cell
culture layout, and the resulting concentration field determines the additional
state-energy increment $\Delta E$ contributed by the indirect effect for each
cell at each time step.

\paragraph{Post-irradiation evolution.}
Using the direct and indirect contributions to $\Delta E$, the cell state
transition process is simulated over a total duration of 1667~minutes.
Each cell is initially assigned one of the five cell-cycle phases (G$_0$, G$_1$,
S, G$_2$, or M) at random. Irradiation occurs only at the initial time point
($t=0$).  
Two simulations are performed: one including only direct radiation effects, and
one including both direct effects and radiation-induced bystander signaling.
The cell-cycle distribution, cell-state distribution, and cell survival fraction
are computed as outputs.  
All simulation parameters are summarized in Table~\ref{tab:1}.

\begin{center}

\begin{longtable}{ |p{1.5cm} | p{4.cm} | p{1.5cm} | p{1.7cm} | p{5cm} | } 
  \hline
 Name & Definition & Value & Unit & Reference \\ 
  \hline
  $X_d$ & Dimension of cell culture in $x$ direction & 5 & mm &   \\ 
  \hline
  $Y_d$ & Dimension of cell culture in $y$ direction & 5 & mm &   \\
  \hline
  $d$ &  Dimension of cell home & 0.05 & mm &  \\
  \hline
  $\Delta T_1$ & Time step for updating cell phase & 60 & second &  \\
  \hline
  $\Delta T_2$ & Time step for updating cell state & 60 & second &  \\
  \hline
  $T_1$ & Duration time of G$_1$ phase & 15 & hour & Adopted from \cite{28}\\
  \hline
  $\sigma_1$ & Standard deviation of $T_1$ & 0.25 & hour & \cite{8} \\
\hline
  $T_2$ & Duration time of S phase & 9 & hour & Adopted from \cite{28}\\
\hline
  $\sigma_2$ & Standard deviation of $T_2$ & 0.25 & hour & Adopted from \cite{28}\\
\hline
  $T_3$ & Duration time of G$_3$ phase & 3 & hour & Adopted from \cite{28}\\
\hline
  $\sigma_3$ & Standard deviation of $T_3$ & 0.25 & hour & Adopted from \cite{28}\\
\hline
  $T_4$ & Duration time of M phase & 1 & hour & Adopted from \cite{28}\\
\hline
  $\sigma_4$ & Standard deviation of $T_4$ & 0.25 & hour & Adopted from \cite{28}\\
\hline
  $\alpha$ & $\alpha$ parameter for DSB number phase & 0.2497 & number & Estimated in this work, detailed estimation process seen in supplementary material\\
\hline
  $\beta$ & $\beta$ parameter for integral number & 18.1517 & ml/pg & Estimated in this work, detailed estimation process seen in supplementary material\\
\hline
  $E_1$ & Mean state energy of S$_1$ state & 0 &  & Basis parameter, taking as zero\\
\hline
  $E_2$ & Mean state energy of S$_2$ & 18.15 &  & Estimated in this work, detailed estimation process seen in supplementary material\\
\hline
  $E_3$ & Mean state energy of S$_3$ & 48.47 &  & Estimated in this work, detailed estimation process seen in supplementary material\\
\hline
  $\sigma$ & Standard deviation of state energy & 6.96 &  & Estimated in this work, detailed estimation process seen in supplementary material\\
\hline
  $f_{G_1}$ & Radiation sensivity factor of G$_1$ phase & 1 &  & Basic parameter, taking as unit\\
\hline
  $f_S$ & Radiation sensivity factor of S phase & 0.816 &  & Extract the radiation sensitivity factors based on these published cell survival fraction curves. Here we take one published cell survival fraction curve in \cite{29}\\
\hline
  $f_{G_2}$ & Radiation sensivity factor of G$_2$ phase & 1.015 &  & Extract the radiation sensitivity factors based on these published cell survival fraction curves. Here we take one published cell survival fraction curve in \cite{29}\\
\hline
  $f_M$ & Radiation sensivity factor of M phase & 1.015 &  & Extract the radiation sensitivity factors based on these published cell survival fraction curves. Here we take one published cell survival fraction curve in \cite{29}\\
\hline
$\lambda_1$ & Fast decay constant of DSB & 3.31 & hour$^{-1}$ & Adopted from \cite{30} \\
\hline
$\lambda_2$ & Slow decay constant of DSB & 0.14 & hour$^{-1}$ & Adopted from \cite{30} \\
\hline
$f_1$ & Weight fraction of fast decay of DSB & 0.62 &  & Adopted from \cite{30} \\
\hline
$f_2$ & Weight fraction of slow decay of DSB & 0.38 &  & Adopted from \cite{30} \\
\hline
$T_t$ & Total simulated time & 1667 & minute & \\
\hline
$N$  & Number of initial seeded cells & 1000 & & \\ 
  \hline
\caption{Simulation parameters using the cellular automaton method for cell state transition.}
\label{tab:1}
\end{longtable}
\end{center}

\section{Results}
\subsection{Cell Phase Evolution}
\begin{figure}[h]
\centering
\includegraphics[width=1.0\textwidth]{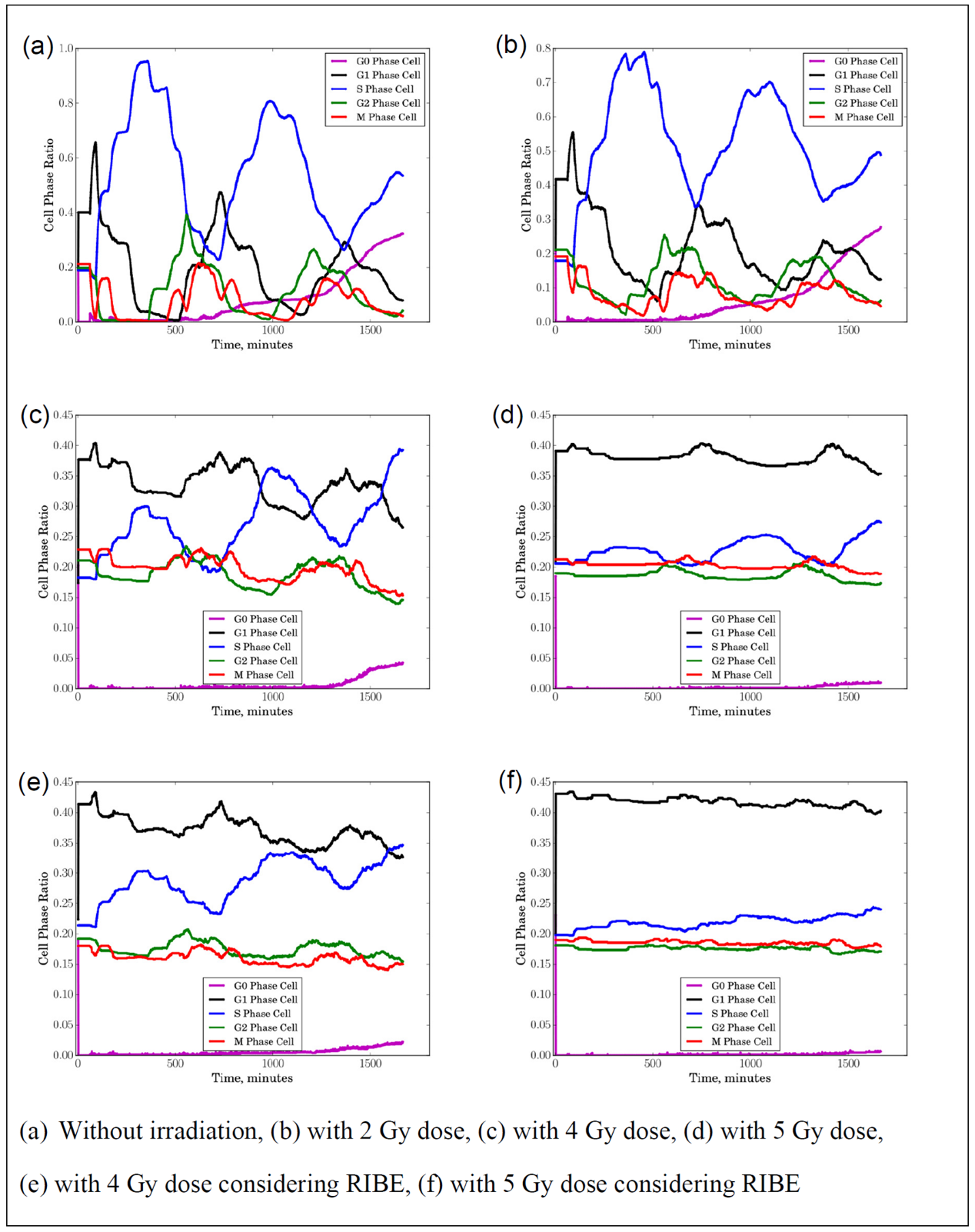}  
\caption{The cell phase distribution of 1000 cells under different dose irradiation.}
\label{fig:5} 
\end{figure}

Figure~\ref{fig:5}a shows the cell-phase distribution of 1000 seeded cells without irradiation. As expected, the cells progress through normal proliferation, and the phase ratios vary with time according to the underlying cell-cycle dynamics. As proliferation proceeds, an increasing number of cells become contact-inhibited, causing more cells to enter and remain in the quiescent G$_0$ phase. The phase ratios of the interphase phases (G$_1$, S, G$_2$) exhibit a periodic pattern corresponding to repeated cell-cycle progression.

Figure~\ref{fig:5}b shows the cell-phase distribution after irradiation with 2~Gy, without including bystander effects. Relative to the unirradiated condition, the fractions of cells in G$_1$, S, and G$_2$ decrease due to radiation-induced cell death. Figure~\ref{fig:5}c presents the results for 4~Gy irradiation, where a substantial reduction in the number of cells progressing to mitosis is observed, and the periodic pattern seen in the interphase phases becomes severely disrupted.

At 5~Gy irradiation (Figure~\ref{fig:5}d), the cell-phase distribution differs markedly from the no-irradiation scenario. Instead of a clear periodic pattern, the phase ratios exhibit much smaller temporal variation due to the greatly reduced number of proliferating cells.

Figure~\ref{fig:5}e illustrates the cell-phase distribution for 4~Gy irradiation when bystander effects are included. A clear difference emerges: fewer cells progress to mitosis compared to the case without bystander signaling. A similar trend is observed at 5~Gy when bystander effects are included (Figure~\ref{fig:5}f), further demonstrating that bystander signaling amplifies radiation-induced suppression of cell-cycle progression.

\subsection{Cell State Evolution}
\begin{figure}[h]
\centering
\includegraphics[width=1.0\textwidth]{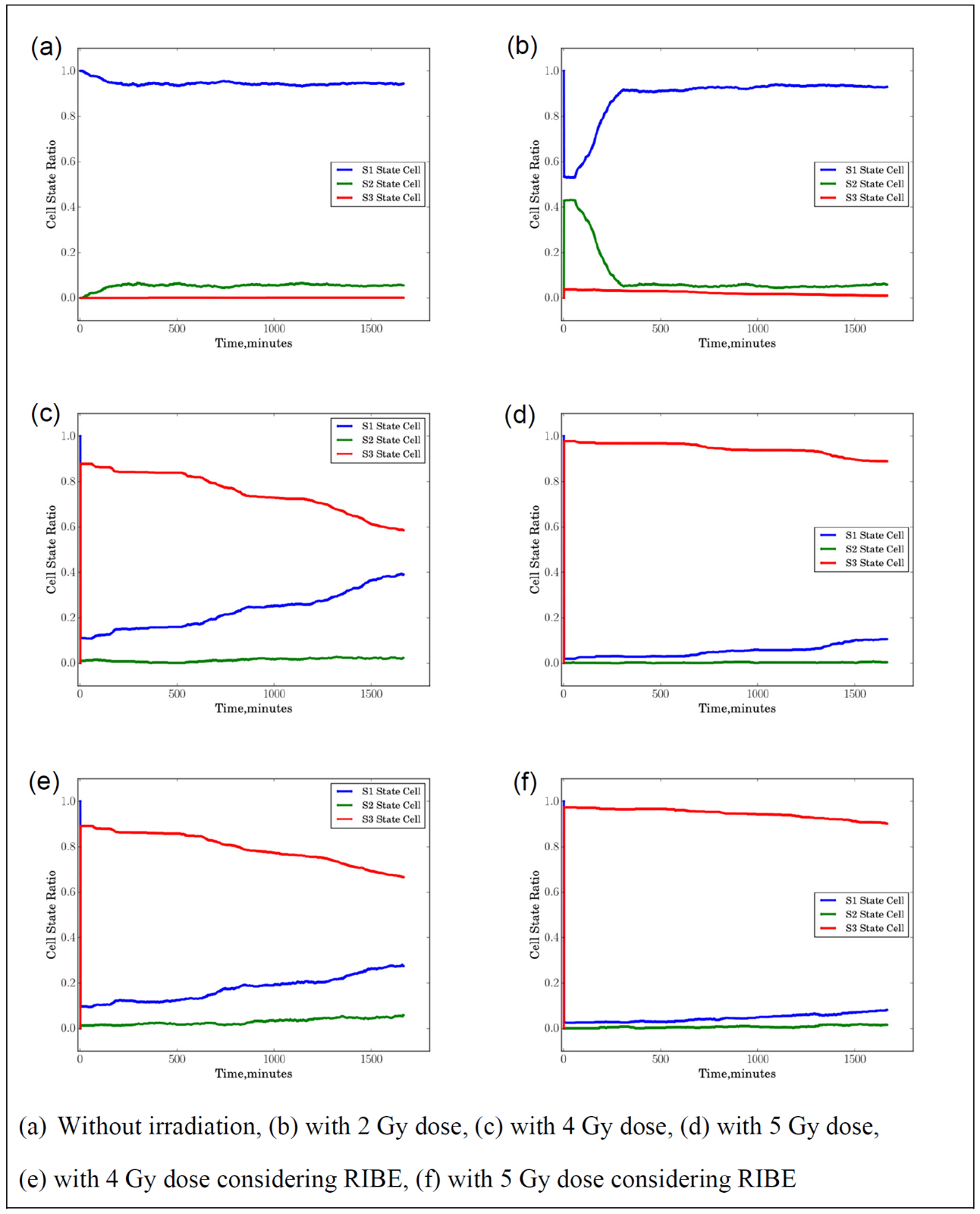}  
\caption{The cell state distribution of 1000 cells under different dose irradiation.}
\label{fig:6} 
\end{figure}

Figure~\ref{fig:6}a shows the cell-state distribution of 1000 seeded cells without irradiation. In the absence of radiation, cells ideally remain in the healthy state $S_1$, so the state distribution should remain nearly constant. However, due to spontaneous death and spontaneous phase arrest, a small fraction of cells enters the arrested state $S_2$ or the death state $S_3$, as reflected in the figure.

Figure~\ref{fig:6}b shows the cell-state distribution following 2~Gy irradiation. After exposure, the fraction of healthy cells ($S_1$) decreases while the fraction of arrested cells ($S_2$) increases. Because 2~Gy induces some cell death but does not eliminate the entire population, $S_1$ gradually recovers over time due to proliferation of surviving cells and partial recovery of sublethally damaged cells. Correspondingly, the fraction of arrested cells initially increases and then slowly decreases as the system stabilizes.

Figure~\ref{fig:6}c presents the cell-state distribution after 4~Gy irradiation. A more pronounced reduction in the $S_1$ population is observed, accompanied by a substantial increase in the death-state fraction ($S_3$). As long as a small population of cells survives the initial irradiation, the healthy-cell fraction increases gradually over time due to proliferation, while the fraction of dead cells decreases correspondingly.

Figure~\ref{fig:6}d shows the results for 5~Gy irradiation. In this case, nearly all cells are lethally damaged shortly after exposure, leading to an overwhelming dominance of $S_3$. The state distribution exhibits minimal temporal variation thereafter, reflecting the very limited capacity for post-irradiation proliferation at this dose level.

Figure~\ref{fig:6}e shows the 4~Gy results when bystander effects are included. The inclusion of bystander signaling results in a modest reduction in the fraction of healthy cells compared to the simulation without bystander effects. However, the magnitude of the difference is relatively small (less than 10\% at all time points). A similar pattern is seen at 5~Gy when bystander effects are included, as shown in Figure~\ref{fig:6}f.

\subsection{Cell Survival Curve}
\begin{figure}[h]
\centering
\includegraphics[width=1.0\textwidth]{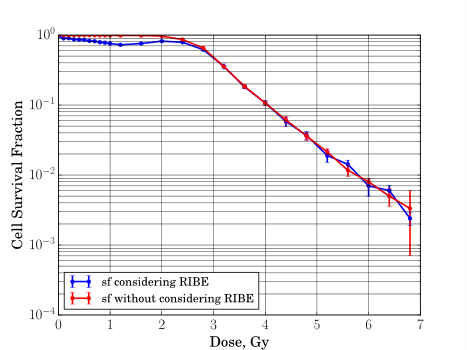}  
\caption{The cell survival fraction curve of 1000 cells in monolayer cell culture irradiated by 1 MeV electrons}
\label{fig:7} 
\end{figure}

We obtained cell survival fraction curves of the seeded cells as shown in Figure \ref{fig:7}. For the simulation case without modeling bystander effects, the cell survival curve captures the basic characteristics of survival fraction curve in experiment. There is a shoulder in the low dose region for the low LET radiation which is 1 MeV electron as simulated in this work, and there is an exponential drop when the dose passes the shoulder region. For the simulation case with modeling bystander effects, we also can observe that there is a hyper-radiosensitivity region when the dose is lower than 2 Gy. 
 
\section{Discussion}

\section*{Discussion}

In this study, we introduced a prototype agent-based simulation module, the \textit{Physical--Bio Translator}, which couples track-structure level energy-deposition information with a stochastic cell-state transition model to simulate post-irradiation cellular responses. Within this framework, each cell is represented by a discrete state (e.g., healthy, arrested, dead) and evolves according to probabilistic transition rules informed by both direct radiation damage and bystander signaling. Using this model, we demonstrated example simulations of cell-cycle evolution, phenotype evolution, and survival in a monolayer cell culture, with and without bystander effects. These results illustrate that the proposed formalism can capture several qualitative features of radiation response in a way that is computationally tractable and compatible with agent-based, multicellular simulations.

This work is motivated by the broader vision, outlined in the Introduction, of developing multiscale digital twins for radiation therapy that link physical energy deposition to biological responses across cellular and tissue scales. Existing radiobiological models---including the linear--quadratic (LQ) model, repair--misrepair models, lethal--potentially lethal models, and two-lesion kinetic models---have made substantial contributions to our understanding of radiation effects and remain central to radiotherapy research and practice.\cite{31,32,33,34,35,36,37} Many of these models have clear microdosimetric foundations and provide mechanistic interpretations at the level of DNA damage and lesion interactions. Our intention is not to replace or dismiss this body of work. Instead, the Physical--Bio Translator is intended as a complementary approach that explores whether a state-based stochastic representation, explicitly linked to track-structure energy deposition, can serve as one building block within a future multiscale digital-twin framework.

An important feature of the present approach is that we did not impose \textit{a priori} assumptions about survival-curve shape or lethal-lesion distributions. Rather, cell survival emerges from the cumulative effect of state transitions driven by the underlying damage-energy inputs. The resulting survival fraction curves exhibit some qualitative behaviors that have been reported experimentally, such as hyper-radiosensitivity-like features at low doses in certain settings.\cite{38,39,40} In the illustrative simulations presented here, the current implementation predicts minimal net biological effect below approximately 2.5~Gy for the specific parameter set used. This behavior should be interpreted with caution and viewed as a property of this particular example rather than as a general prediction of the framework. In practice, low-dose responses are known to vary substantially across cell lines and experimental conditions, and the model is not intended, in its present form, to capture all possible survival-curve behaviors. A rigorous comparison with experimental datasets will be required to determine whether a given parameterization can reproduce observed responses for a specific biological system. In future work, we plan to calibrate model parameters against measured survival data and to investigate whether additional mechanisms (e.g., low-dose hypersensitivity, adaptive responses, or more detailed checkpoint dynamics) are needed to reproduce dose--response relationships across the full clinically relevant range.

The broader aim of developing this model is to provide a flexible, mechanistically motivated tool for linking radiation energy deposition to probabilistic cellular outcomes, rather than to claim a fully unified or superior description of radiation biology. The Physical--Bio Translator has several features that make it attractive for integration into multiscale simulations. First, the model parameters (e.g., those governing state energies, transition thresholds, and bystander dynamics) are defined with explicit biophysical or phenomenological meaning and, in principle, can be constrained by experiments. Second, the modular structure of the model allows additional mechanisms---such as immune interactions, specific signaling pathways, or heterogeneity in cell radiosensitivity---to be incorporated without fundamentally altering the underlying formalism. In this sense, the current work should be viewed as a proof-of-concept step toward more comprehensive digital-twin components, rather than a complete description of radiation response.

Several limitations of the present study merit emphasis. The bystander effect is modeled using a simplified reaction--diffusion representation and translated into state-energy increments, which cannot fully capture the rich diversity of intercellular signaling pathways observed experimentally. Experimental validation will therefore be essential. For example, microbeam irradiation of single cells or small clusters, as well as conventional linac-based irradiation of cultured cells, could be used to compare model predictions with measured outcomes such as clonogenic survival, DNA damage repair kinetics, or temporal patterns of cell-cycle arrest. Such studies would both test the assumptions embedded in the Physical--Bio Translator and help refine its parameters.

Another important limitation is the current lack of formal uncertainty quantification. Because the model contains parameters that are not yet tightly constrained by data, predictions should be interpreted qualitatively. Incorporating uncertainty quantification---for example through Bayesian calibration, sensitivity analysis, or ensemble simulations---will be critical for assessing the robustness of model predictions and for identifying which parameters have the greatest impact on observable outcomes. Finally, the simulations presented here are restricted to a relatively simple monolayer cell culture and do not include tissue-level heterogeneity, vascular structure, immune components, or long-term adaptive responses. These features will be essential for any clinically meaningful digital twin of radiation therapy.

In summary, the Physical--Bio Translator provides an initial demonstration that a cell-state transition model, coupled to track-structure energy deposition and bystander signaling, can be implemented in an agent-based framework and used to generate qualitatively reasonable patterns of cell-cycle dynamics, phenotype distributions, and survival. The present work should be viewed as an exploratory step toward the broader goal of building multiscale, mechanistic digital twins for radiation therapy, rather than as a replacement for existing radiobiological models. Continued refinement, experimental validation, and rigorous parameter calibration will be required to determine the extent to which this framework can quantitatively reproduce known biological responses and contribute meaningfully to predictive modeling in radiation oncology.

\section{Conclusions}
In this study, we developed and implemented a mechanistic, cell-state–based framework for quantifying radiation-induced cellular effects and demonstrated its use within an agent-based simulation environment. We introduced a new mathematical formalism for modeling cell-state transitions after irradiation and showed how it can be coupled to track-structure–level energy deposition and bystander signaling to generate emergent patterns of cell-cycle evolution, phenotype distribution, and survival. The resulting Physical–Bio Translator should be viewed as a proof-of-concept module that provides a tractable way to link physical dose deposition to probabilistic cellular outcomes, rather than as a complete description of radiation response. Nevertheless, this framework represents a key step toward integrating cell-scale mechanistic models into multiscale digital-twin architectures for radiation therapy. With further refinement, experimental calibration, and extension to tissue- and patient-level processes, the concepts developed here could contribute to future digital twin systems aimed at improving the personalization and biological realism of radiation treatment planning.
\section{Acknowledgments}
RL acknowledges support from the Department of Radiation Oncology at the University of Nebraska Medical Center through faculty startup funding. Additional funding for this project was provided, in part, by The Otis Glebe Medical Research Foundation.

\end{document}